# High Speed Data Transmission over GI-MMF Using Mode Group Division Multiplexing

Bernd Franz, Liaquat Ali, Laurent Schmalen, Wilfried Idler

*Abstract* — We transmitted 4x28 Gbit/s over 5-km standard GI-MMF using four different mode groups in combination with direct detection and on-off-keying format (OOK). Due to the square-law detection in the receiver MIMO processing is not effective. Therefore proper mode group selective multiplexing and de-multiplexing is essential.

*Index Terms* — Graded-index multi-mode fiber, Optical fiber communication, Spatial multiplexing, Direct detection, On-off-Keying

## I. INTRODUCTION

SPATIAL multiplexing using standard graded-index multi-mode fibers (GI-MMF) in combination with On-Off-Keying (OOK) modulation and direct detection might be an economical solution for short haul transmission systems, e.g. for data center interconnects or in the local metro region. Since the effectiveness of MIMO processing is limited due to square-law detection, cross-talk between different spatial transmission paths should be minimized.

The key elements for spatial multiplexing systems are spatial multiplexer and de-multiplexer. Different approaches for these devices have been published [1,2] but up to now only methods based on free space Fourier optics show the ability to produce the desired modes with high suppression of other unwanted modes for a high number of modes. But even here it must be recognized that this comes at the expense of high attenuation [3] or a certain portion of unwanted modes that have to be accepted [4].

If a high transmission bandwidth is needed, differential mode delay (DMD) should be reduced for each transmit channel. GI-MMF with a parabolic index allows the propagation of mode groups which comprise modes with approximately equal phase constants and group delays, respectively [5]. Linear polarized modes $LP_{\nu\mu}$ are determined by their azimuthal order $\nu$ and radial order $\mu$. All modes with the mode number m fulfilling the equation $m = \nu + 2\mu + 1$ belong to such a mode group of order m (MGm). For example the mode group MG5 with m=5 includes the modes LP02, LP21a and LP21b. These mode groups have the potential to provide a transmission channel with high bandwidth [4].

We report on the transmission of 4x28-Gbit/s OOK modulated data signals over a span of 5 km of standard GI-MMF using mode group division multiplexing (MGDM) in combination with direct detection.

## II. EXPERIMENTAL SET-UP

Fig. 1 shows the experimental set-up. A commercially available programmable pattern generator delivers a $2^{15}$-1 PRBS at a bit rate of 28 Gbit/s. It is followed by a Mach-Zehnder modulator modulating the incoming light in OOK format. The wavelength of the light is 1550 nm. The data signal is then distributed to a subsystem of power splitters, EDFAs and delay lines so that four temporally de-correlated copies of the signal are fed to the spatial 4:1 multiplexer [1].

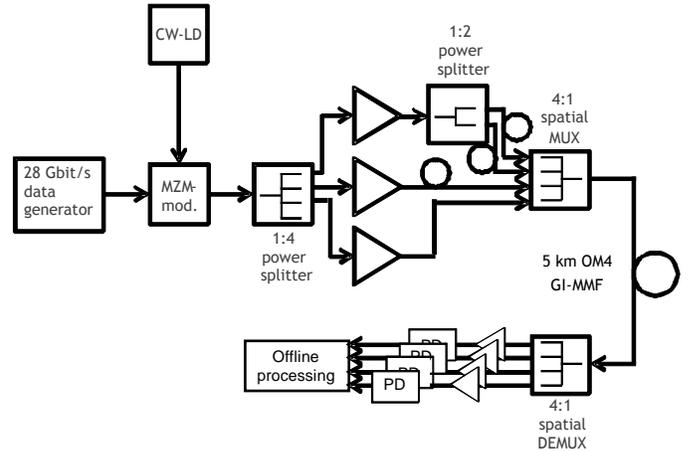

Fig. 1: Block diagram of the experimental set-up

This device converts the LP01 modes coming from the four single mode fibers (SMF) into modes belonging to different mode groups and feeds them into the MMF, see Fig. 2. An optical beam combiner combines the four differently diffracted beams. A 5-km long standard graded-index MMF serves as transmission medium. The spatial 1:4 de-multiplexer located at the receiver is identical to the 4:1 multiplexer of Fig. 2, where in- and outputs are inverted.

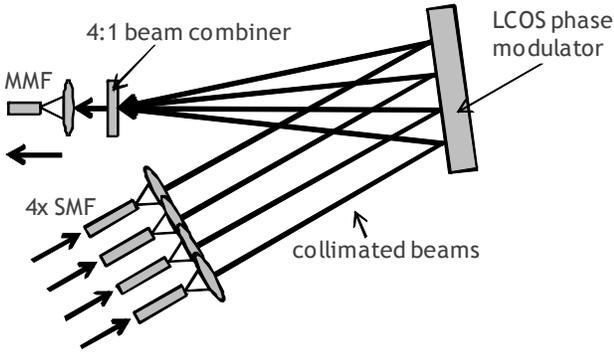

Fig. 2: Principle of the spatial 4:1 multiplexer

While the transmitter needs to excite only one mode per mode group, the receiver must detect all available modes of the desired mode group. The reason for this is that the modes within a certain mode group couple with each other during the propagation along the fiber because they have approximately identical phase constants. For experimental simplicity, and without loss of generality, we used mode groups MG3 to MG6 which include fewer modes than the mode groups with higher order and detected only the strongest mode of each received mode group.

The mix of incoming modes is diffracted by the spatial modulator so that the desired mode is converted to the LP01 mode of the following SMF. These SMFs serve as modal filters so that only the desired signal is fed to the following photo diodes after amplification by EDFAs. Here the optical signals are directly detected and sampled by a real-time sampling scope. The four simultaneously sampled signals have then been processed offline.

## III. Experimental Results

In a first experiment we transmitted only one 28 Gbit/s data signal using in turn the mode groups MG3 to MG6 as transmission channel. The sampling scope was set to its maximum sampling rate of 80 GSa/s, which is only available if one input is used. Fig. 3 shows the eye diagrams of the received signals. All eye diagrams show clear openings. We sampled the received data signal in twenty data sequences which contain 1048576 samples each. After arithmetical re-sampling and synchronization with the transmitted data we obtained 327670 bits per data sequence. We arithmetically evaluated the bit error rates (BER) for the received signals using all twenty data sequences. All four mode groups transmitted the data without errors.

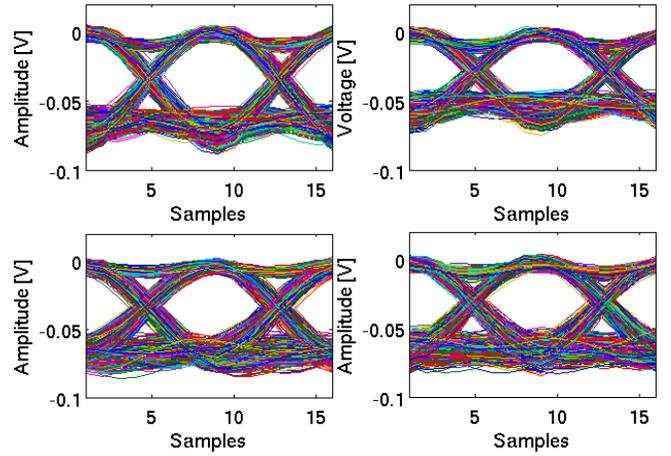

Fig. 3: Eye diagrams of the received 28 Gbit/s data signals using MG3 (top left), MG4 (top right), MG5 (bottom left) and MG6 (bottom right)

In the next experiment we evaluated the performance of the link if four mode groups have been simultaneously used as independent channels. Each channel transmitted a 28 Gbit/s data signal resulting in an overall bit rate of 112 Gbit/s. These four data signals have been simultaneously detected, sampled and then processed in the same way as described in the first experiment. Since the sampling scope was here limited to a sampling rate of 40 GSa/s we obtained 688107 bits per data sequence. We applied off-line re-sampling and synchronization. During these measurements we observed a temporally varying cross-talk between the received signals. In order to catch the impact of the time-variant cross-talk we recorded in parallel thirty temporally successive data sequences for each of the four detected modes and calculated the bit error rate (BER) for each data sequence. As expected we obtained different BER for different data sequences according to the fluctuations of the cross-talk. However, the measurements showed that the cross-talk can be seen as time-invariant during the recording of one data sequence. This can be seen in Fig. 4 which shows as an example the temporal error distribution during two successively recorded data sequences from the channel using mode group MG4. Transmitted and received bits were compared and a one marks an error. A uniform error distribution can be seen for both data sequences. On the other hand these plots show also the strong BER variation between two data sequences of the same channel.

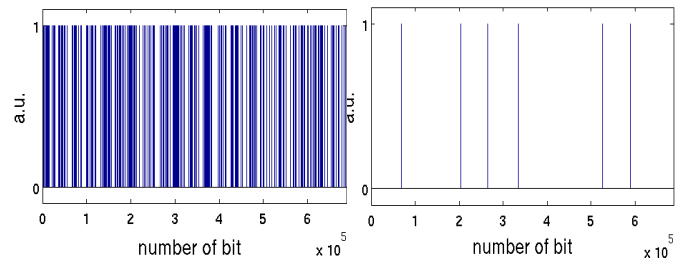

Fig. 4: Temporal error distribution of two succeeding data sequences of the same mode group

Fig. 5 underlines this behavior. It shows the variations of the detected errors per recorded data sequence for all recorded sequences of the channel using mode group MG4. The BER ranges from below $1.5*10^{-6}$ (no error detected) to $1*10^{-3}$ (674 errors detected).

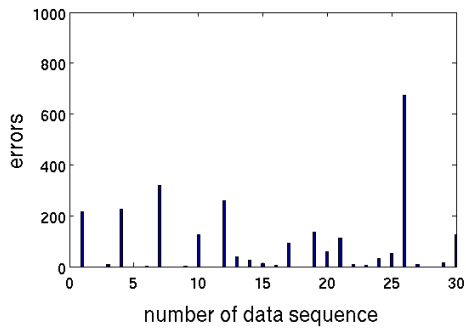

Fig. 5: Error distribution over the recorded data sequences for the channel using mode group MG4

Fig. 6 illustrates the frequency of the different BERs in four histograms for the four different channels. The insets indicate the modes which belong to the respective channel.

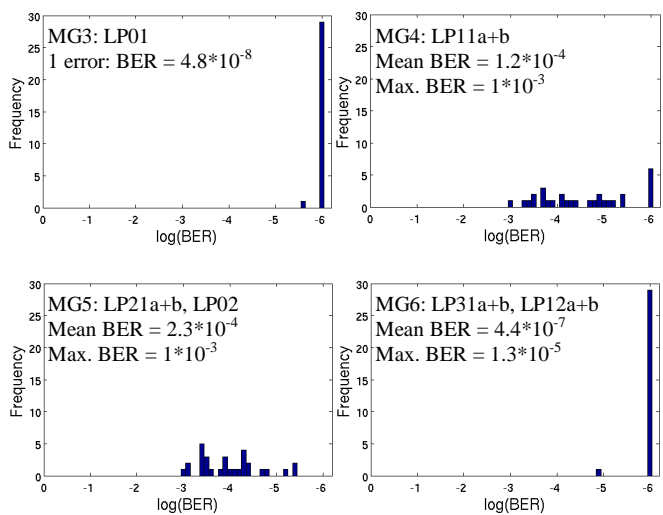

Fig. 6: Frequency distribution of the BER for the transmission channels using MG3, MG4, MG5 and MG6

The transmission quality strongly varies between the different channels. During our experiments we did not manage to obtain the temporal stability of our spatial multiplexers and de-multiplexers which is required for negligible cross-talk at all four channels at the same time. The result presented here shows the lowest BER values we obtained during a series of received data sequences. The BER values vary between $1*10^{-3}$ and below $1.4*10^{-6}$ (1 error) for all four channels depending on the cross-talk. The average BER calculated from 30 measurements amounts to $4.8*10^{-8}$ (1 error) for mode group MG3, $1.2*10^{-4}$ for mode group MG4, $2.3*10^{-4}$ for mode group MG5 and $4.4*10^{-7}$ for mode group MG6. All these BERs can be corrected by a Reed-Solomon-FEC with 12% overhead to below $10^{-12}$. The net bit rate is then reduced to 100 Gbit/s.

## IV. CONCLUSIONS

We showed the feasibility of a 112 Gbit/s transmission over 5-km standard GI-MMF using four different mode groups. Although a time-variant cross-talk inhibits an error-free transmission the maximum observed BER of below $1*10^{-3}$ can be corrected by a Reed-Solomon-FEC with 12% overhead. The net bit rate would then be reduced to 100 Gbit/s. However, the single channel experiment shows that a transmission with very low BER is possible for all four channels if the crosstalk between the mode groups would be low enough.

The experimental set-up suffered from technological limitations, e.g. the precision of the used micromanipulators. An integrated solution with diffractive elements should substantially relax the mechanical requirements so that the crosstalk will be reduced.

REFERENCES

[1] B. Franz et al., "Mode Group Division Multiplexing in Graded-Index Multimode Fibers", Bell Labs Technical Journal, Vol. 18, Issue 3, pp. 153–172, Dec. 2013.
[2] N. K. Fontaine et al., "Mode-Selective Dissimilar Fiber Photonic-Lantern Spatial Multiplexers for Few-Mode Fiber", ECOC 2013, paper PD1.C.3.
[3] J. Carpenter et al., "Precise modal excitation in multimode fibre for control of modal dispersion and mode-group division multiplexing", ECOC 2011, paper We.10.P1.62.
[4] B. Franz et al., "Experimental Evaluation of Principal Mode Groups as High Speed Transmission Channels in Spatial Multiplex Systems", Photonics Technology Letters, Vol. 24, Issue 16, pp. 1363-1365 (2012).
[5] R. Olshansky, "Propagation in glass optical waveguides", Reviews of Modern Physics, Vol. 51, No. 2, pp. 341-367, Apr. 1979.